\documentclass[11pt,a4paper]{article}

\usepackage[utf8]{inputenc}
\usepackage[english]{babel}

\usepackage[top=20mm,bottom=30mm,left=20mm,right=20mm]{geometry}

\usepackage{authblk}
\usepackage{hyperref}       
\usepackage{url}            
\usepackage{amsfonts}      
\usepackage{nicefrac}       
\usepackage{microtype}      
\usepackage{amsmath}
\usepackage{amssymb}
\usepackage{physics}
\usepackage{graphicx}
\usepackage{subcaption}
\usepackage[export]{adjustbox}
\usepackage{wrapfig}
\usepackage{array}
\usepackage{cite}
\newcolumntype{M}[1]{>{\centering\arraybackslash}m{#1}}
\graphicspath{ {./images/} }

\author[1]{S. I. Godunov}
\author[2,3]{E. K. Karkaryan}
\author[1]{V. A. Novikov}
\author[2,4]{A. N. Rozanov}
\author[1]{M. I. Vysotsky}
\author[1]{E.~V.~Zhemchugov~\thanks{Corresponding author. E-mail: evgenii.zhemchugov@cern.ch}}

\affil[1]{I.E. Tamm Department of Theoretical Physics, Lebedev Physical Institute, 119991 Moscow, Russia}
\affil[2]{Institute for Theoretical and Experimental Physics, 117218 Moscow, Russia}
\affil[3]{Moscow Institute of Physics and Technology (State University), 141701 Moscow, Russia}
\affil[4]{Centre de Physique des Particules de Marseille, CPPM, Aix-Marseille
Universite, CNRS/IN2P3, F-13288 Marseille, France
}

\title{LHC as a photon-photon collider: bounds on $\Gamma_{X\rightarrow\gamma\gamma}$}

\begin{document}
\maketitle
\begin{abstract}
  In the relatively recent CMS data, there is a hint on the existence
  of a resonance with the mass 28 GeV decaying to a $\mu^+ \mu^-$ pair and
  produced in association with a $b$ quark jet and a second jet. Such a
  resonance should also couple to photons through the fermion loop, therefore it
  can be searched for in ultraperipheral collisions (UPC) of protons. We set an
  upper bound on the $X \gamma \gamma$ coupling constant from the data on $\mu^+
  \mu^-$ pair production in UPC at the LHC. Our approach can be used for similar
  resonances should they appear in the future.
\end{abstract}

\section{Introduction}
LHC designed as a proton-proton collider can also be considered as a
photon-photon collider in which photons are produced in
ultraperipheral collisions of protons. The interest in studying
$\gamma\gamma$ collisions is twofold: first, QED processes like
$\gamma\gamma\rightarrow l^+l^-$ \cite{paper1,doppaper1,doppaper2},
$\gamma\gamma\rightarrow W^+W^-$
\cite{paper2,doppaper3,doppaper4,doppaper5},
$\gamma\gamma\rightarrow\gamma\gamma$
\cite{paper3,doppaper6,doppaper7} are investigated at very high
energies never before accessible at particle accelerators, and second,
production of new exotic particles can be looked for. The case of
long-lived heavy charged particles was considered in
\cite{doppaper8}. Dark matter particles are discussed in
\cite{doppaper9,doppaper10,doppaper11}. In the paper \cite{paper4} the
production of exclusive $\gamma\gamma\rightarrow\mu^+\mu^-$ events in
proton-proton collisions at a center-of-mass energy of
$13\; \text{TeV}$ with the ATLAS detector was analyzed. The
measurement was performed in the dimuon invariant mass interval
$12\: \text{\text{GeV}}<m_{\mu^+\mu^-}<70\: \text{\text{GeV}}$. If a
resonance with the mass in this interval does exist and can decay to a
$\mu^+\mu^-$ pair, we can obtain an upper bound on its coupling with
two photons from the data provided in \cite{paper4}. A hint of such
a resonance $X$ with the mass
$\left(28.3\pm 0.4\right)\: \text{\text{GeV}}$ was reported by the CMS
Collaboration \cite{paper5}, and in what follows, we will obtain bounds
on its coupling to two photons. However, being universal, our approach
can be used for another resonance if it exists.\footnote{
 In particular, the $X$ production mechanism in inclusive $pp$ collisions is not
 straightforward and requires introduction of other new particles~\cite{paper6}.
 This is not relevant for the $X$ production in $\gamma \gamma$ collision.
}

As it was noticed in \cite{paper6}, $X$ can be responsible for the
deviation of the measured value of the muon anomalous magnetic moment
$a_{\mu}$ from its theoretical value. Introducing the coupling $Y$ of
the scalar $X$ to muons according to
\begin{equation}
  \label{eq1}
  \Delta \mathcal{L}=Y\overline{\mu}\mu X,
\end{equation}
it was obtained in \cite{paper6} that for $Y=0.041\pm0.006$ one loop
contribution $\delta a^X_{\mu}=(29\pm 8)\times 10^{-10}$ explains the
deviation of the measured value of $a_{\mu}$ from the Standard Model
result. It was also shown that such couplings are consistent with
other experimental bounds.

With this value of $Y$, we get
\begin{equation}
  \label{eq2} 
  \Gamma_{X\rightarrow\mu^+\mu^-}=\frac{Y^2}{8\pi}M_X
  \bigg(1-\frac{4m^2_{\mu}}{M^2_X}\bigg)^{3/2}=(1.8\pm0.5)\:
  \text{MeV},
\end{equation}
while according to \cite{paper5} the width of the peak is
\begin{equation}
  \label{eq3}
  \Gamma^{\rm exp}_X=(1.8\pm0.8)\: \text{\text{GeV}},
\end{equation}
which is close to the detector mass resolution for a dimuon system
$\sigma=0.45\: \text{GeV}$. That is why we will also consider the case of
$\Gamma_X$ approximately equal to $\Gamma_{X\rightarrow\mu^+\mu^-}$ given in
(\ref{eq2}).

\section{The fiducial cross section of the
  $pp(\gamma\gamma)\rightarrow pp\mu^+\mu^-$ reaction}

We are interested in the contribution of the $X$ resonance to this
cross section. In \cite{paper4}, the cross section of $\mu^+\mu^-$
production was measured in four intervals of the muon pair invariant mass
on which the entire interval
$12\: \text{GeV}<m_{\mu^+\mu^-}<70\: \text{GeV}$ was divided. We are
interested in the interval
$22\: \text{GeV}<m_{\mu^+\mu^-}<30\: \text{GeV}$, for which, according
to Table\: 3 of \cite{paper4},
\begin{equation}
  \label{eq4}
  \frac{d\sigma^{\rm exp}}{d m_{\mu^+\mu^-}}=(0.076\pm0.005)\:
  \frac{\text{pb}}{\text{GeV}},\quad \text{hence}\quad
  \sigma^{\rm exp}=(0.61\pm0.04)\: \text{pb}.
\end{equation}

This cross section measurement corresponds to the fiducial region
$p^{\mu}_T>\hat{p}_T=6\: \text{GeV}$ and $|\eta|<\hat{\eta}=2.4$,
where $p^{\mu}_T$ is the component of the muon momentum transversal to
the proton beam and $\eta$ is the muon pseudorapidity:
$\eta=-\ln{\tan({\theta/2})}$, where $\theta$ is the angle between the
muon momentum and the beam. The ATLAS muon spectrometer is measuring
muon momentum up to $\abs{\eta}=2.7$, but the trigger chambers cover
the range $\abs{\eta}<2.4$ that corresponds to the pseudorapidity
cutoff given above.

According to the equivalent photons approximation the cross section of
$\mu^+\mu^-$ pair production in ultraperipheral collisions is given
by
\begin{equation}
  \label{eq***}
  \sigma(pp(\gamma\gamma)\rightarrow pp\mu^+\mu^-)=
  \int\limits^{\infty}_0 d\omega_1\int\limits^{\infty}_0 d\omega_2
  \sigma(\gamma\gamma\rightarrow\mu^+\mu^-)n(\omega_1)n(\omega_2),
\end{equation}
where $n(\omega)$ is the equivalent photons spectrum. In the leading
logarithmic approximation (LL),
\begin{equation}
  \label{eq100}
  n(\omega) \approx n_{\rm LL}(\omega) =
  \frac{2 \alpha}{\pi \omega} \ln \frac{\hat{q} \gamma}{\omega},
\end{equation}
where $\alpha$ is the fine structure constant,
$\gamma = 6.93 \times 10^3$ is the Lorentz factor of the proton with
the energy 6.5 TeV, and $\hat q$ is the maximal photon momentum at
which the proton does not disintegrate. In this approximation the
integrals in Eq.~(\ref{eq***}) are divergent, and the integration
domain is cut off explicitly with $\hat{q}\gamma$,
\begin{equation}
  \label{eq*}
  \sigma_{\rm LL}(pp(\gamma\gamma)\rightarrow pp\mu^+\mu^-) =
  \int\limits^{\hat{q} \gamma}_{m^2_{\mu}/\hat{q} \gamma} d\omega_1
  \int\limits^{\hat{q} \gamma}_{m^2_{\mu}/\omega_1} d\omega_2
  \sigma(\gamma\gamma\rightarrow\mu^+\mu^-)n_{\rm LL}(\omega_1)n_{\rm LL}(\omega_2),
\end{equation}
The value of $\hat q$ is determined by the proton form factor and
numerically $\hat{q}\approx0.20\: \text{GeV}$ \cite{paper7}.

It is convenient to substitute the integration over photon energies by
integration over $s=4\omega_1\omega_2$ and
$x={\omega_1}/{\omega_2}$. Then Eq.~(\ref{eq***}) changes to
\begin{equation}
  \label{eq6}
  \sigma(pp(\gamma\gamma)\rightarrow pp\mu^+\mu^-) =
  \int\limits^{\infty}_{(2m_{\mu})^2}\sigma(\gamma\gamma\rightarrow\mu^+\mu^-)ds
  \int\limits^{\infty}_0\frac{dx}{8x}n\bigg(\sqrt{\frac{sx}{4}}\bigg)
  n\bigg(\sqrt{\frac{s}{4x}}\bigg).
\end{equation}

To take the experimental cuts into account, we substitute
$\sigma(\gamma\gamma\rightarrow\mu^+\mu^-)$ by the differential over
$p_T$ cross section,
\begin{equation}
  \label{eq7}
  \sigma(pp(\gamma\gamma)\rightarrow pp\mu^+\mu^-) =
  \int\limits^{\infty}_{(2m_{\mu})^2} ds
  \int\limits^{\sqrt{s}/2}_0\frac{d\sigma(\gamma\gamma\rightarrow\mu^+\mu^-)}{dp_T}dp_T
  \int\limits^{\infty}_0 \frac{dx}{8x}n\bigg(\sqrt{\frac{sx}{4}}\bigg)
  n\bigg(\sqrt{\frac{s}{4x}}\bigg).
\end{equation}
It is then straightforward to implement cuts over $s$ and $p_T$ by
changing the integration limits to
$\hat{s}_{\rm min} < s < \hat{s}_{\rm max}$ and
$\hat{p}_T < p_T < \sqrt{s} / 2$ [assuming
$\hat{s}_{\rm min} \geqslant (2 \hat p_T)^2 \gg (2m_\mu)^2$]. To implement the
cutoff over pseudorapidity, one should integrate over $x$ in the
interval \cite{paper7},
\begin{equation}
  \label{eq8}
  \frac{1}{\hat{x}}<x<\hat{x},\quad \text{where}\:
  \hat{x}=\exp(2\hat{\eta})\frac{1-\sqrt{1-4p^2_T/s}}{1+\sqrt{1-4p^2_T/s}}.
\end{equation} 

Let us note that in the leading logarithmic approximation from the
condition $\omega\lesssim \hat{q}\gamma$, it follows that $x$ should be
always smaller than $(2\hat{q}\gamma/\sqrt{s})^2$. For numerical
values of $\hat{\eta}$, $\hat{p}_T$, and
$\hat{s}=\left\{\hat{s}_{\rm min},\hat{s}_{\rm max}\right\}$, we are
interested in, and for $x$ from the interval (\ref{eq8}), this demand is
satisfied.

Thus, for the fiducial cross section we obtain
\begin{equation}
  \label{eq9}
  \sigma^{\hat{s},\hat{p}_T, \hat{\eta}}_{\rm fid} =
  \int\limits^{\hat{s}_{\rm max}}_{\hat{s}_{\rm min}}ds
  \int\limits^{\sqrt{s}/2}_{\hat{p}_T}\frac{d\sigma(\gamma\gamma\rightarrow\mu^+\mu^-)}{dp_T}dp_T
  \int\limits^{\hat{x}}_{1/\hat{x}}\frac{dx}{8x}
  n\bigg(\sqrt{\frac{sx}{4}}\bigg)
  n\bigg(\sqrt{\frac{s}{4x}}\bigg),
\end{equation}   
where $\hat{x}$ is defined in (\ref{eq8}). In the leading logarithmic
approximation, the fiducial cross section is
\begin{equation}
  \label{eq**}
  \sigma^{\hat{s},\hat{p}_T, \hat{\eta}}_{\rm fid, LL} =
  \frac{\alpha^2}{\pi^2}\int\limits^{\hat{s}_{\rm max}}_{\hat{s}_{\rm min}}
  \ln^2{\frac{(2\hat{q}\gamma)^2}{s}}\frac{ds}{s}
  \int\limits^{\sqrt{s}/2}_{\hat{p}_T}\frac{d\sigma(\gamma\gamma\rightarrow\mu^+\mu^-)}{dp_T}
  \left[1-\frac{1}{3}\left(\frac{\ln{\hat{x}}}{\ln{\frac{(2\hat{q}\gamma)^2}{s}}}\right)^2\right]\ln{\hat{x}}\: dp_T.
\end{equation}
 
\begin{figure}[t]

  \begin{subfigure}{0.33\textwidth}
    \centering
    \includegraphics{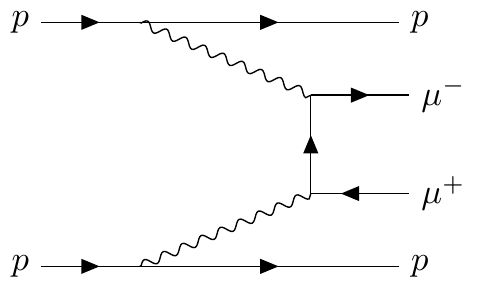}
    \caption{}
    \label{fig:subim1}
  \end{subfigure}
  \begin{subfigure}{0.33\textwidth}
    \centering
    \includegraphics{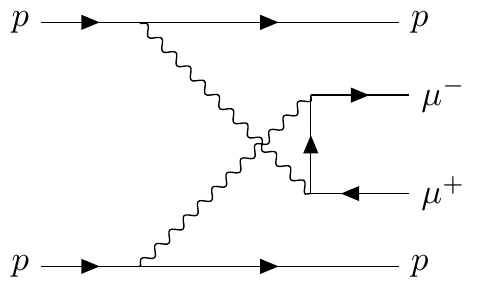}
    \caption{}
    \label{fig:subim2}
  \end{subfigure}
  \begin{subfigure}{0.33\textwidth}
    \centering
    \includegraphics{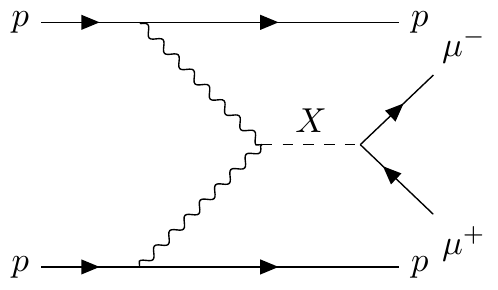}
    \caption{}
    \label{fig:subim3}
  \end{subfigure}

  \caption{Diagrams which contribute to the production of muon pair in
    ultraperipheral $pp$ collisions}
  \label{fig:image2}
\end{figure}

Let us begin with the calculation of the Standard Model contribution
to the cross section of $\mu^+\mu^-$ pair production, given by the
diagrams shown in Figs.~\ref{fig:subim1},~\ref{fig:subim2}.

The expression for the differential cross section is 
\cite[\S 88]{paper8}
\begin{equation}
  \label{eq10}
  d\sigma(\gamma\gamma\rightarrow\mu^+\mu^-) =
  \frac{2\pi\alpha^2}{s^2}\bigg(\frac{s+t}{t}+\frac{t}{s+t}\bigg)dt =
  \frac{8\pi\alpha^2}{sp_T}\frac{1-2p^2_T/s}{\sqrt{1-4p^2_T/s}}dp_T.
\end{equation}

Substituting it in (\ref{eq**}) and integrating over $p_T$, we get
\begin{align}
  \label{eq11}
  \sigma^{\hat{s},\hat{p}_T, \hat{\eta}}_{\rm fid, LL}&\approx
  \frac{8\alpha^4}{\pi}\int\limits^{\hat{s}_{\rm max}}_{\hat{s}_{\rm min}}
  \ln^2\frac{(2\hat{q}\gamma)^2}{s}\frac{ds}{s^2}
  \left\{
  \hat{\eta}\left[\ln(\frac{1+\sqrt{1+4\hat{p}^2_T/s}}{1-\sqrt{1-4\hat{p}^2_T/s}})-\sqrt{1-\frac{4\hat{p}^2_T}{s}}\right]-
  \right.\nonumber\\ &\left.
  -\frac{1}{4}\ln^2\left(\frac{1+\sqrt{1+4\hat{p}^2_T/s}}{1-\sqrt{1-4\hat{p}^2_T/s}}\right)
  +\frac{1}{2}\sqrt{1-\frac{4\hat{p}^2_T}{s}}\ln(\frac{1+\sqrt{1+4\hat{p}^2_T/s}}{1-\sqrt{1-4\hat{p}^2_T/s}})
  \right\}
  =0.73\: \text{pb},
\end{align}
where we neglected the small second term in the square brackets in
(\ref{eq**}) in order to perform integration analytically. Taking into
account the omitted term and integrating numerically in (\ref{eq**}),
instead of $0.73\; \text{pb}$, we obtain $0.68\; \text{pb}$.

More accurate calculation depends on the internal structure of proton and the
probability for the protons to survive the collision. For the latter, we will
use the expression suggested in Ref.~\cite{doppaper12},
\begin{equation}
  \label{eq13.1}
  P(b) = \left( 1 - \mathrm{e}^{-\frac{b^2}{2 B}} \right)^2,
\end{equation}
where $b$ is the impact parameter of the collision, and $B$ was
measured to be $19.7~\text{GeV}^{-2}$ in the case of $pp$ collisions
with the energy 7 TeV \cite{doppaper13}. To utilize this function, we
introduce the equivalent photon spectrum at the distance $b$ from the
source particle $n(b, \omega)$, such that
\begin{equation}
  \label{eq13.2}
  n(\omega) = \int n(b, \omega) d^2 b.
\end{equation}

Then the leading logarithmic spectrum \cite[\S 15.5]{paper13},
\begin{equation}
  n_{\rm LL}(b,\omega)=\frac{\alpha\omega}{\pi^2\gamma^2}K_1\left(\frac{b\omega}{\gamma}\right),
\end{equation}
where $K_1$ is the modified Bessel function of the second kind (the
Macdonald function).

In the framework of the parton model and following Ref.~\cite{Cahn-Jackson},
Eq.~(\ref{eq***}) is replaced with
\begin{align}
  \label{eq13.4}
  \sigma(pp (\gamma \gamma) \to pp \mu^+ \mu^-)
  = \int\limits_0^\infty d \omega_1
  \int\limits_0^\infty d \omega_2&
  \sigma(\gamma \gamma \to \mu^+ \mu^-)\times\nonumber\\
  &\times\int d^2 b_1
  \int d^2 b_2
  n(b_1, \omega_1)
  n(b_2, \omega_2)
  P(\lvert \mathbf{b}_1 - \mathbf{b}_2 \rvert).
\end{align}

This change is then propagated into Eq.~(\ref{eq9}),
\begin{align}
  \label{eq13.5}
  \sigma_{\rm fid}^{\hat s, \hat p_T, \hat \eta}
  = \int\limits_{\hat{s}_{\rm min}}^{\hat{s}_{\rm max}} ds&
  \int\limits_{\hat p_T}^{\sqrt{s} / 2} d p_T
  \frac{d \sigma(\gamma \gamma \to \mu^+ \mu^-)}{d p_T}\times\nonumber\\
  &\times\int\limits_{1/\hat x}^{\hat x}
  \frac{dx}{8x}
  \int\limits_{b_1>0} d^2 b_1
  \int\limits_{b_2>0} d^2 b_2
  \, n \left(b_1, \sqrt{\frac{sx}{4}} \right)
  \, n \left(b_2, \sqrt{\frac{s}{4x}} \right)
  P(\lvert \mathbf{b}_1 - \mathbf{b}_2 \rvert).
\end{align}

The internal structure of proton is characterized by the Dirac form
factor \cite{paper11}
\begin{equation}
  \label{eq12}
  F_1(Q^2) = G_D(Q^2)\left[ 1 + \frac{(\mu_p - 1) \tau}{1 +
      \tau} \right],\quad G_{D}\left(Q^{2}\right)=\frac{1}{(1+\frac{Q^2}{\Lambda^2})^2},
\end{equation}
where $Q^{2}=-q^2$, $q$ is the photon 4-momentum,
$\tau = Q^2 / 4 m_p^2$, $m_{p}$ is the proton mass, and
$\mu_{p}=2.7928473508(85)$ is the proton magnetic
moment~\cite{codata2014}, $G_D(Q^2)$ is the dipole form factor with
$\Lambda$ being strictly fixed by the proton charge radius:
$\Lambda^2 = {12}/{r_{p}^{2}}$, $r_{p}=0.8751(61)$
fm~\cite{codata2014}.

The form factor enters Eqs. (\ref{eq***}), (\ref{eq13.5}) through the
equivalent photon spectrum \cite{paper7},
\begin{equation}
  n(\omega) = \frac{\alpha}{\pi^2\omega}
    \int
      \frac{
        \vec q_\perp^{\; 2} \, F_{1}^2(\vec q_\perp^{\; 2} + \omega^2 / \gamma^2)
      }{
        (\vec q_\perp^{\; 2} + \omega^2 / \gamma^2)^2
      }
    \, \mathrm{d}^2 q_\perp,
  \label{spectrum-ff}
\end{equation}
\begin{equation}
  n(b, \omega)
  = \frac{\alpha}{\pi^2 \omega}
     \left[
       \int \mathrm{d} q_\perp
         q_\perp^2
         \frac{F_{1}(q_\perp^2 + \omega^2 / \gamma^2)}
              {q_\perp^2 + \omega^2 / \gamma^2}
        J_1(b q_\perp)
    \right]^2,
 \label{survival-spectrum}
\end{equation}
where $\vec q_\perp$ is the photon transversal momentum, $J_1$ is the Bessel
function of the first kind.

The so-called survival factor
$S^2_{\gamma\gamma}$\cite{paper7,doppaper12,paper10} is defined as the
ratio of the integrands in
Eqs.~(\ref{eq***}),~(\ref{eq13.4}),\footnote{Survival factor can be
  also defined in a more elaborate way: on the amplitude
  level. See~\cite{Khoze:2000db,Khoze:2002dc,Khoze:2014aca,
    Harland-Lang:2014lxa,Harland-Lang:2015cta,Khoze:2017sdd,
    Harland-Lang:2020veo} for details.

  Let us also note that definition of $S_{\gamma \gamma}^2$ in
  Ref.~\cite{paper9} [Eq.~(7)] is different: Ref.~\cite{paper9}
  requires that the new system is produced outside of the colliding
  particles, while Ref.~\cite{paper7} imposes no such restriction. The
  latter is more accurate when the new particles do not interact
  strongly, so we use the Ref.~\cite{paper7} definition of
  $S_{\gamma \gamma}^2$ here. In paper~\cite{Harland-Lang:2020veo}, it
  was specifically stressed that impact parameter cut like
  in~Ref.~\cite{paper9} is unphysical.}
\begin{equation}
  S^2_{\gamma\gamma}=\frac{\int_{b_1>0}
    \int_{b_2>0}n(b_1,\omega_1)n(b_2,\omega_2)P(|\mathbf{b_1-b_2}|)d^2b_1d^2b_2}
  {n(\omega_1)n(\omega_2)},
\end{equation}
however, in this paper, it is not calculated explicitly;
Eq. (\ref{eq13.5}) is used instead.

Calculations for each interval of muon pair invariant mass for
Eq.~(\ref{eq**}), Eq.~(\ref{eq9}) with the
spectrum~(\ref{spectrum-ff}), and Eq.~(\ref{eq13.5}) with the
spectrum~(\ref{survival-spectrum}) are presented in the
Table~\ref{table}.  One can see that accounting for inelastic $pp$
scattering reduces the theoretical result by 5\% approximately.

\begin{table}[t]
\centering
\caption{The measured cross section for each interval of muon pair
  invariant mass and the corresponding theoretical calculations with
  different approximations: Eq.~(\ref{eq**}) is the calculation with
  the equivalent photon spectrum taken in the leading logarithmic
  approximation; Eqs.~(\ref{eq9}), (\ref{spectrum-ff}) is the
  calculation taking into account the proton electromagnetic form
  factor; Eqs.~(\ref{eq13.5}), (\ref{survival-spectrum}) also accounts
  for the probability of strong interactions at small impact
  parameters. "Survival ratio" is the ratio of the preceding two
  columns. Note that for the interval 30--70 GeV the cutoff
  $\hat p_T = 10$~GeV as it is in \cite{paper4}.  }
\label{table}
\begin{tabular}{|M{2cm}|M{2.3cm}|M{2.4cm}|M{2.4cm}|M{2.4cm}|M{1.5cm}|}
 \hline
  $m_{\mu^+\mu^-},\: \text{GeV}$ & $\sigma^{\rm exp},\: \text{pb}$~\cite{paper4}
  & Leading logarithmic approx., \:
    Eq.~(\ref{eq**})
  & With the form factor, \:
    Eqs.~(\ref{eq9}), (\ref{spectrum-ff})
  & Also with the survival factor, \:
    Eqs.~(\ref{eq13.5}), (\ref{survival-spectrum})
  & Survival ratio\\
 \hline
 12--17 & $1.22\pm0.07$ & $1.25$ & $1.28$  & $1.24$  & $0.970$\\
 17--22 & $0.82\pm0.05$ & $0.87$ & $0.896$ & $0.866$ & $0.967$\\
 22--30 & $0.61\pm0.04$ & $0.68$ & $0.703$ & $0.677$ & $0.963$\\
 30--70 & $0.52\pm0.04$ & $0.49$ & $0.506$ & $0.483$ & $0.953$\\
\hline
\end{tabular}
\end{table}

The amplitude of the $\mu^+\mu^-$ pair production through intermediate
$X$ boson in $\gamma\gamma$ collisions [see Fig.~\ref{fig:subim3}] is
given by the following expression:
\begin{equation}
  \label{eq14}
  A=\kappa F^1_{\mu\nu}F^2_{\mu\nu}\frac{1}{s-M^2_X+i\Gamma_XM_X}\overline{\mu}\mu Y, 
\end{equation}
where $\kappa$ is the $X\gamma\gamma$ coupling constant so that
$\Gamma_{X\to\gamma\gamma}=(\kappa^2M^3_X)/(16\pi)$.  For the cross
section of the $\gamma\gamma\rightarrow X\rightarrow\mu^+\mu^-$
reaction, we obtain
\begin{equation}
  \label{eq15}
  |A|^2=\kappa^2Y^2M^6_X\frac{1}{(s-M^2_X)^2+\Gamma^2_XM^2_X},
\end{equation}
\begin{equation}
  \label{eq16}
  \sigma_{\gamma\gamma\rightarrow X \rightarrow \mu^+\mu^-} =
  \frac{2\pi}{M^2_X}\frac{\Gamma_{X\rightarrow
      \gamma\gamma}\Gamma_{X\rightarrow \mu^+\mu^-}}
  {(\sqrt{s}-M_X)^2+\Gamma^2_X/4},
\end{equation}
where the factor $2$ takes into account identity of photons.

In the limit $m_\mu \to 0$ chiralities of the muons produced through
the diagrams in Figs.~\ref{fig:subim1},~\ref{fig:subim2} are not the
same as in Fig.~\ref{fig:subim3}. Consequently, these diagrams do not
interfere in this limit. Even with nonzero $m_\mu$ the interference is
zero at $s = M_X^2$ because then the phase between the sum of the
diagrams in Figs.~\ref{fig:subim1},~\ref{fig:subim2} and the diagram
in Fig.~\ref{fig:subim3} is $\pi / 2$. For other values of $s$, the
interference is suppressed relatively to $X$ contribution by the
factor
$\frac{\alpha \Gamma_{X}}{\sqrt{\Gamma_{X \to \mu^+ \mu^-} \Gamma_{X
      \to \gamma \gamma}}}\frac{m_\mu}{M_{X}} \left( 1 -
  \frac{M_X^2}{s} \right)$, which is less than $10^{-2}$ for the
largest allowed values of $\Gamma_{X\rightarrow \gamma \gamma}$ in
both cases of the narrow or the wide resonance ($\Gamma_X = 1.8$~MeV
or $1.8$~GeV, respectively).

In order to impose the cut on the transverse momentum of muons with
the help of expression (\ref{eq9}) the following differential cross
section is used:
\begin{equation}
  \label{eq17}
  d\sigma=\frac{|A|^2}{32\pi s}\frac{d(4p^2_T/s)}{\sqrt{1-4p^2_T/s}}.
\end{equation}

Substituting (\ref{eq17}) and (\ref{eq15}) in (\ref{eq**}) and
performing integration over $p_T$, we obtain
\begin{multline}
  \label{eq18}
  \sigma^{\hat{s},\hat{p}_T, \hat{\eta}}_{\rm fid, LL}(X) =
  \frac{8\alpha^2\Gamma_{X\rightarrow\gamma\gamma}\Gamma_{X\rightarrow\mu^+\mu^-}}
  {\pi M^2_X}
  \int\limits^{\hat{s}_{\rm max}}_{\hat{s}_{\rm min}}
  \frac{ds}{(s-M^2_X)^2+\Gamma^2_X M^2_X}
  \ln^2\frac{(2\hat{q}\gamma)^2}{s}
\times \\
    \times  \left[\sqrt{1-\frac{4\hat{p}^2_T}{s}}
    \left(2\hat{\eta}+\ln(\frac{1-\sqrt{1-4\hat{p}^2_T/s}}{1+\sqrt{1-4\hat{p}^2_T/s}})\right)
    -\ln{\frac{4\hat{p}^2_T}{s}}\right].
\end{multline}

In the case of a narrow resonance
$\Gamma_X\approx\Gamma_{X\rightarrow\mu^+\mu^-}=(1.8\pm0.5)\:
\text{MeV}$, the integration can be performed analytically, and we
obtain
\begin{multline}
  \label{eq19}
  \sigma^{\hat{s},\hat{p}_T, \hat{\eta}}_{\rm fid, LL}(X) =
  \frac{8\alpha^2\Gamma_{X\rightarrow\gamma\gamma}\Gamma_{X\rightarrow\mu^+\mu^-}}
  {\Gamma_X M^3_X}
  \ln^2\frac{(2\hat{q}\gamma)^2}{M^2_X}\times
  \\\times\left[\sqrt{1-\frac{4\hat{p}^2_T}{M^2_X}}\left(2\hat{\eta}
      +\ln(\frac{1-\sqrt{1-4\hat{p}^2_T/M^2_X}}{1+\sqrt{1-4\hat{p}^2_T/M^2_X}})\right)
    -\ln{\frac{4\hat{p}^2_T}{M^2_X}}\right]\approx\\
    \approx 6.1\times 10^4 \frac{\Gamma_{X\rightarrow\mu^+\mu^-}}{M_X}\frac{\Gamma_{X\rightarrow\gamma\gamma}}{\Gamma_X} \: \text{pb}.
\end{multline}

However, if $\Gamma_X=(1.8\pm0.8)\: \text{GeV}$, then the width of the
resonance almost equals $\sqrt{\hat{s}_{\rm max}}-M_X=2\; \text{GeV}$,
so the integration should be done numerically, and we obtain
\begin{equation}
  \label{eq20}
  \sigma^{\hat{s},\hat{p}_T, \hat{\eta}}_{\rm fid, LL}(X)\approx
  49\: \frac{\Gamma_{X\rightarrow\gamma\gamma}}{M_X}\: \text{pb}.
\end{equation}

\section{Numerical estimates}
From the third line of the second and the fifth columns of the
Table~\ref{table}, we see that the contribution of the resonance $X$ into the
fiducial cross section of muon pair production is bounded in the following
way:\footnote{
  When calculating the upper limit, in the case of negative signal,
  Ref.~\cite{Cowan:2010js} suggests using zero instead of the negative value.
  This makes the upper limit a little less strong. This approach is widely used
  in the LHC experimental community, so we follow it here.
}
\begin{equation}
  \label{21}
  \sigma_{\rm fid}(X)\lesssim 0.10 \: \text{pb} \; \text{at 99.5\% confidence level.}
\end{equation}

Comparing this number with the expression (\ref{eq19}), we get that if
$\Gamma_X\approx\Gamma_{X\rightarrow\mu^+\mu^-}$ then the upper bound
on $\Gamma_{X\rightarrow\gamma\gamma}$ is
\begin{equation}
  \label{eq22}
  \text{Br}(X\rightarrow\gamma\gamma)<2.6\times 10^{-2},\:
  \Gamma_{X\rightarrow\gamma\gamma}< 46~\text{keV} \approx 1.6\times
  10^{-6}\: M_X \;
  \text{at 99.5\% confidence level}\; (\Gamma_X=1.8\: \text{MeV}).
\end{equation}

If the width of $X$ is given by (\ref{eq3}) then the bound extracted
from Eq.(\ref{eq20}) is
\begin{equation}
  \label{eq23}
  \text{Br}(X\rightarrow\gamma\gamma)<3.2\times 10^{-2},\;
  \Gamma_{X\rightarrow\gamma\gamma}< 58~\text{MeV} \approx 2\times 10^{-3}\;
  M_X\;
  \text{at 99.5\% confidence level}\; (\Gamma_X=1.8\: \text{GeV}).
\end{equation}

\begin{figure}[t]
  \centering
  \includegraphics{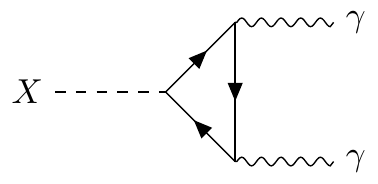}
  \caption{Coupling of $X$ to two photons through a fermion loop}
  \label{fig:Xgammagamma}
\end{figure}

Resonance $X$ couples with photons through a triangle diagram with
fermion running in the loop (see Fig.~\ref{fig:Xgammagamma}). Let us
check that the corresponding decay probability does not violate bounds
just obtained.

The amplitude generated by the triangle diagram with a fermion $f$
running in the loop equals \cite{paper12}
\begin{equation}
  \label{eq24}
  A=\frac{\alpha F}{4\pi}Y_{Xff}\frac{1}{m_{f
    }}XF^1_{\mu\nu}F^2_{\mu\nu}.
\end{equation}

The width equals
\begin{equation}
  \label{eq25}
  \Gamma_{X\rightarrow\gamma\gamma} =
  \frac{\alpha^2 F^2}{256\pi^3}Y^2_{Xff}\Big(\frac{M_X}{m_f}\Big)^2 M_X,
\end{equation}
where 
\begin{align}
  \label{eq26}
  F&=-2\beta[(1-\beta){\varkappa}^2+1],\; \beta=\frac{4m^2_f}{M^2_X},\\
  \varkappa
   &\label{eq27}
     =
  \begin{cases}
    \arctan(\frac{1}{\sqrt{\beta-1}}), \; \beta>1\\
    \frac{1}{2}\Big[i\ln(\frac{1+\sqrt{1-\beta}}{1-\sqrt{1-\beta}})+\pi\Big], \; \beta<1.
  \end{cases}
\end{align}
For $m_f\ll M_X$, we obtain $F\sim(m_f/M_X)^2$, and for $m_f\gg M_X$, we
obtain $F\rightarrow-4/3$.

In the case of muon running in the loop, we get
$\Gamma_{X\rightarrow\gamma\gamma}\approx 10^{-11} M_X$, which is much
smaller than bounds (\ref{eq22}), (\ref{eq23}). For a hypothetical
fermion with a mass much larger than $M_X$, the width is also very
small. However, for $m_f\sim M_X$ and $Y_{Xff}\sim 1$, it approaches
keV:
$\Gamma_{X\rightarrow\gamma\gamma}(m_f=M_X/2)\approx3Y^2_{Xff}\:\text{keV}$.

\section{Conclusions}

A scalar resonance with the mass 28 GeV coupling to muons in the way
consistent with the recent CMS data~\cite{paper5} is also consistent
with the measurements of the cross section for muon pair production in
ultraperipheral collisions at the LHC~\cite{paper4} provided that the
width of its decay to a pair of photons
$\Gamma_{X \to \gamma \gamma} < 46$~keV or 58~MeV depending on whether
the width $\Gamma_X = 1.8$~GeV reported in Ref.~\cite{paper5} is the
real width of the resonance or an artifact of the detector mass
resolution.

The difference between the leading logarithmic approximation and the
calculation that takes into account both the proton form factor and
the survival factor for the protons colliding with the energy 13~TeV
is at the level of few percent. Integration of the logarithmic
approximation can often be performed analytically while the form
factor and especially the survival factor require computationally
expensive numerical calculations. Therefore, cross sections for
ultraperipheral collisions of protons in the lower region of invariant
masses of the produced system can be estimated in the logarithmic
approximation with the form factor and the survival factor taken into
account as needed.

Our study demonstrates that we can look for New Physics in
ultraperipheral collisions at the LHC.

We are grateful to V.B. Gavrilov and A.N. Nikitenko, who have brought
the CMS observation of $X(28\: \text{GeV})$ to our attention. We are
grateful to V.A. Khoze for drawing our attention to
papers~\cite{Khoze:2000db,Khoze:2002dc,Khoze:2014aca,
  Harland-Lang:2014lxa,Harland-Lang:2015cta,Khoze:2017sdd,
  Harland-Lang:2020veo}. We are supported by the Russian Science
Foundation Grant No. 19-12-00123.

\bibliographystyle{unsrt}

\end{document}